\documentclass[twocolumn,showpacs,preprintnumbers,aps,prl,lengthcheck]{revtex4-1}
\usepackage[english]{babel}
\usepackage{amssymb}
\usepackage{amsmath}
\usepackage{graphicx}
\usepackage{dcolumn}
\usepackage{bm}
\usepackage{hyperref}
\usepackage{makeidx}
\usepackage{placeins}
\usepackage[usenames,dvipsnames]{color}

\makeindex

\begin{document}

\title{Conformal perturbation of off-critical correlators in the 3D Ising universality class}
\author{M. Caselle$^{1}$, G. Costagliola $^{1}$, N. Magnoli$^{2}$}
   \affiliation{
$^1$ Dipartimento di Fisica, Universit\`a di Torino and INFN, Via P. Giuria 1,
10125, Torino,
Italy.\\
$^2$ Dipartimento di Fisica, Universit\`a di Genova and INFN, Via Dodecaneso 33, 16146,
Genova, Italy.
}
\date{\today}

\begin{abstract}
Thanks to the impressive progress of conformal bootstrap methods we have now very precise estimates 
of both scaling dimensions and OPE coefficients for 
several 3D universality classes. We show how to use this information to obtain similarly precise 
estimates for off-critical correlators using conformal perturbation.
We discuss in particular the $\langle \sigma (r) \sigma (0) \rangle$, $\langle \epsilon (r) \epsilon (0) \rangle$ and $\langle
\sigma (r) \epsilon (0) \rangle$ two point functions in the high and low temperature regimes of the 3D 
Ising model and evaluate the leading and next to leading terms in the $s = t r^{\Delta_{t}}$ 
expansion, where $t$ is the reduced temperature. Our results for $\langle \sigma (r) \sigma (0) \rangle$ agree 
both with Monte Carlo simulations 
and with a set of experimental estimates of the critical scattering function.
\end{abstract}

\maketitle

\section{Introduction}
Our understanding of critical phenomena in three dimensions has been greatly improved in the past 
few years by the remarkable progress of the Conformal Bootstrap program.
In particular, using Conformal Bootstrap techniques, very precise estimates of scaling dimensions 
and Operator Product Expansion (OPE) coefficients could be obtained for several three dimensional  
models 
\cite{ElShowk:2012ht,El-Showk:2014dwa,Gliozzi:2013ysa,Gliozzi:2014jsa,Kos:2015mba,Komargodski:2016auf,Kos:2016ysd} . 
Among them a particular attention was devoted to the 3D Ising model, which plays somehow a 
benchmark role in this context, since in this case both high precision Monte Carlo results
\cite{hasen,Caselle:2015csa,Costagliola:2015ier} and accurate experimental estimates \cite{exp} 
exist for these quantities.

The aim of this letter is to leverage the accurate knowledge of the model that we have at the 
critical point to predict the behaviour of off-critical correlators in the whole scaling region 
using Conformal Perturbation Theory (CPT).
We shall discuss, as an example, the thermal perturbation of the $\langle \sigma (r) \sigma (0) \rangle$, $\langle
\epsilon (r) \epsilon (0) \rangle$ and $\langle \sigma (r) \epsilon (0) \rangle$ correlators in the 3D Ising model 
both above and below the critical temperature, but our results are of general validity and may be 
applied to any model for which scaling dimensions and critical OPE constants are known with 
sufficient precision. In this sense our work is a natural extension of the Conformal Bootstrap 
Approach.

It is important to stress the role played by Conformal Symmetry  in our analysis. In fact  
the general form of correlation functions and in particular their dependence on the expansion 
parameter (in our case the reduced temperature $t$) was already understood more than  fifty years 
ago \cite{Fisher1968,Brezin:1974zz} and can be easily obtained using standard scaling arguments valid 
in general for any critical point. 
The additional bonus we have in the case of a conformally invariant critical point is that the terms 
appearing in the CPT expansion can be related to the derivatives of the Wilson coefficients, 
calculated at the critical point 
\cite{Guida:1995kc,Guida:1996nm,Caselle:2001zd,Caselle:1999mg}. 
In several cases these derivatives can be evaluated exactly and only depend on the critical 
indices and OPE coefficients of the underlying Conformal Field Theory (CFT). Unfortunately the 
power of this approach is limited by the fact that the vacuum expectation values (VEVs) of the 
relevant operators which appear in the expansion must be known to all orders in the perturbation 
parameter. This means in practical applications that they appear in the CPT expansion as external 
inputs, they are not universal and in general cannot be fixed by using CFT data.
In some cases (in particular in two dimensions and for integrable perturbations) these VEVs can be 
evaluated analytically \cite{Guida:1996nm,Caselle:2001zd,Caselle:1999mg} or numerically using the 
truncated conformal space method \cite{Guida:1997fs}. When this is not possible they must be 
evaluated using independent non-perturbative methods, like strong coupling expansions or Monte Carlo 
simulations. 
There are however exceptions to this rule. The terms involving derivatives of the Wilson 
coefficients of the type $\partial _t C^1 _{O O}$ do not require any external information, they can 
be evaluated exactly (we shall see below an example) and represent thus true, universal, testable 
predictions of the CPT approach. In this sense they are more informative than usual universal 
amplitudes ratios since they test the presence of the whole conformal symmetry in the critical 
theory and not only of scale invariance, that is, instead, a sufficient requirement to construct 
ordinary universal amplitude ratios.

The main goal of this paper is exactly to perform this non trivial "conformality test" in the case 
of the 3D Ising model. As a side result of this analysis we shall be able to give a very precise 
prediction for the scaling behaviour of the $\langle \sigma (r) \sigma (0) \rangle$ correlator in the scaling 
region which we shall successfully compare both with Monte Carlo results and with a set of 
experimental estimates of the critical scattering function for a ${\bf CO_2}$ sample at critical 
density.

\section{Thermal perturbation theory}

Let us consider the thermal perturbation of the 3D Ising 
model, characterized by two relevant operators, the magnetization 
 $\sigma$ and the energy $\epsilon$, whose dimensions are
$\Delta_{\sigma} = 0.5181489 (10) $ and $\Delta_{\epsilon} = 1.412625 (10) $ respectively 
\cite{Komargodski:2016auf,Kos:2016ysd}. 

The perturbed action is given by the 
conformal point action $S_{cft}$, plus a term proportional to  the energy operator:
\begin{equation}\label{act}
S = S_{cft} + t \int  \epsilon (r) \; d{\bf r}
\end{equation}
where $t$ is a parameter related in the continuum limit to the deviation from the critical 
temperature. We shall denote in the following with $\langle ... \rangle _t$ expectation values with respect to the perturbed action $S$ and with $\langle ... \rangle _0$ those with respect to 
the unperturbed conformal invariant action $S_{cft}$.

In \cite{Guida:1996nm} the correlators of two generic operators $O_i$ and $O_j$ of the perturbed CFT were expressed, by using the operator product expansion, in terms 
of the  Wilson coefficients, calculated outside of the critical point:

\begin{equation}
\langle O_i (r) O_j (0) \rangle _t = \sum_{k} C^{k}_{ij}(t,r) \langle O_{k} (0) \rangle_t .
\end{equation}

In order to perform the CPT expansion one has to expand in a Taylor series the Wilson coefficients, while the VEV's 
must be determined in a non perturbative way.
The first few terms of this CPT expansion read

\begin{equation}
\langle O_i (r) O_j (0) \rangle _t = \sum_{k} [C^{k}_{ij}(0,r)+\partial _t C^{k}_{ij}(0,r)+....] \langle O_{k} (0) \rangle_t
\end{equation}

where $\partial _t C^{k}_{ij}(0,r)$ denotes the derivatives of the Wilson coefficients with respect
to $t$ evaluated at the critical point.

\noindent
In  \cite{Guida:1995kc}  it was shown how to calculate these derivatives to any order $n$ and it was proved that they are infrared finite.



Defining $\Delta_{t}=3-\Delta_{\epsilon}$, the perturbed 
one-point functions are (see \cite{Pelissetto:2000ek} for definitions and further information on the scaling behaviour of the model):
\begin{equation}
\langle \epsilon \rangle_t = A^{\pm} |t|^{\frac{\Delta_{\epsilon}}{\Delta_{t}}} \quad
 \langle \sigma \rangle_t = B_{\sigma} (-t)^{\frac{\Delta_{\sigma}}{\Delta_{t}}} .
 \end{equation}

We have the following expression for the first three orders of perturbed two-point function of $\sigma$:

\begin{multline}\label{corrW}
\langle \sigma (r) \sigma (0) \rangle _t = 
C_{\sigma\sigma}^{1} (0,r) \\
+ C_{\sigma\sigma}^{\epsilon} (0,r) A^{\pm} |t|^{\frac{\Delta_{\epsilon}}{\Delta_{t}}}
+ t  \partial_{t} C_{\sigma\sigma}^{1} (0,r) + ... 
\end{multline}

To make contact with the usual definition for the structure constants we factorize the $r$ dependence in the Wilson coefficients:

$$C_{\sigma\sigma}^{1}(0,r) = \frac{1}{r^{2 \Delta_ \sigma}},~~~~~  C_{\sigma\sigma}^{\epsilon}(0,r) = C_{\sigma\sigma}^{\epsilon} r^{\Delta _\epsilon -2 \Delta _\sigma}$$ 
where we have chosen the usual normalization $C_{\sigma\sigma}^{1}=1$ and we know from \cite{Komargodski:2016auf,Kos:2016ysd} that $C^{\epsilon}_{\sigma \sigma} = 1.0518537(41)$.

Following \cite{Guida:1995kc} we can write the derivatives of the Wilson coefficient as:

\begin{multline}\label{wilsder}
\partial _t C^1 _{\sigma  \sigma } (0,r) \\
= -\int (\langle \sigma (r) \sigma (0) \epsilon( r_1 ) \rangle _0  - C^{\epsilon}_{\sigma \sigma} \langle \epsilon({
r_1}) \epsilon(0) \rangle _0 ) d{\bf r_1},
\end{multline}

The three point function reads:
\begin{equation}
\langle \sigma (r) \sigma (0) \epsilon( r_1) \rangle _0  = \frac{C^{\epsilon}_{\sigma \sigma}}{r^{2\Delta_{\sigma}-\Delta_{\epsilon}} r_1 ^{ \Delta _\epsilon } (r^2+r_1^2-2 r r_1~ \cos 
\theta )^{\frac{\Delta_ \epsilon }{2 }}}\nonumber
\end{equation}
while $\langle \epsilon (r) \epsilon (0) \rangle _0 = \frac{1}{r^{2 \Delta_ \epsilon }}$.
and the second term in the integral acts as an infrared counterterm.
The integral (\ref{wilsder}) can be calculated using a Mellin transform technique (see 
\cite{Guida:1996ux}) or numerically.
In the first case only the first term gives a contribution:
\begin{equation}\label{derwilscoeff}
\partial _t C^1 _{\sigma  \sigma } (0,r) = 
r^{\Delta _t - 2 \Delta _\sigma} C^{\epsilon}_{\sigma\sigma}
 \int \frac{1}{y^{\Delta \epsilon}} 
 \frac{1}{(1+y^2-2 y cos \theta )
 ^{\frac{\Delta _\epsilon }{2} }}d {\bf y} \nonumber
\end{equation}
where $y = \frac{r_1}{r}$.
It is convenient to define $ \partial _t C^1 _{\sigma  \sigma } (0,r) \equiv  r^{\Delta _t - 2 \Delta _\sigma} C^{\epsilon}_{\sigma\sigma} I$,
so after performing the angular integrals we get:

\begin{equation}
I = 2 \pi \int \frac{\left(1+ y\right)^{-\Delta _\epsilon +2}-\left(1-y\right)^{-\Delta _\epsilon +2}}{-\Delta _\epsilon +2} 
y^{-\Delta _\epsilon +1} dy\nonumber
\end{equation}
which can be solved in terms of Gamma functions giving as a final result: $ I = -62.5336$.

\noindent
Now we can write, introducing the scaling variable  $s = t r^{\Delta_{t}}$,
the following expression for the perturbed two-point function:
\begin{equation}\label{corr1}
r^{2\Delta_{\sigma}} \langle  \sigma (r) \sigma (0)\rangle _t = 1 + 
C_{\sigma\sigma}^{\epsilon} A^{\pm} |s|^{\frac{\Delta_{\epsilon}}{\Delta_{t}}} -
C_{\sigma\sigma}^{\epsilon} I s 
\end{equation}
Proceeding in the same way we have for the others correlators of interest:

\begin{equation}\label{corr2}
r^{2\Delta_{\epsilon}} < \epsilon (r) \epsilon (0) > _t= 1 + 
C_{\epsilon\epsilon}^{\epsilon} A^{\pm} |s|^{\frac{\Delta_{\epsilon}}{\Delta_{t}}} -
C_{\epsilon\epsilon}^{\epsilon} I s + ... 
\end{equation}

\begin{equation}\label{corr3}
r^{\Delta_{\epsilon}+\Delta_{\sigma}} < \sigma (r) \epsilon (0) >_t =
C_{\sigma\epsilon}^{\sigma} B_{\sigma} |s|^{\frac{\Delta_{\sigma}}{\Delta_{t}}} +
...
\end{equation}

Recalling the value of $C^{\epsilon}_{\sigma \sigma}$ and that
$C^{\epsilon}_{\epsilon \epsilon} = 1.532435(19)$  \cite{Komargodski:2016auf,Kos:2016ysd}  we finally obtain:

$$-C_{\sigma\sigma}^{\epsilon} I =  65.7762...~~~~ -C_{\epsilon\epsilon}^{\epsilon} I  = 
95.8287... $$

This is the main result of our paper. It is interesting to compare these quantities with the second 
terms in the CPT expansions. 
To estimate these terms we need the values of the $A^{\pm}$ which can be easily obtained from the 
very precise estimates of their lattice values reported in \cite{hasen}
$A_{lat}^{+} = -8.572(4)$, $A_{lat}^{-}= 15.987(3)$ and the lattice to continuum conversion 
constants 
$R_{\epsilon} = 0.2377 (9)$, $R_{\sigma} = 0.550 (4)$ \cite{Caselle:2015csa,Costagliola:2015ier}.
We find $A^{+} =-48.7(3)$ and $A^{-}=90.9(6) $ to which correspond the following values for the
terms which appear in the CPT expansion:
\begin{equation}
C_{\sigma\sigma}^{\epsilon} A^{+}= -51.2(3),~~ C_{\sigma\sigma}^{\epsilon} A^{-}= 95.6(6)
\nonumber
\end{equation}

\begin{equation}
C_{\epsilon\epsilon}^{\epsilon} A^{+}=-74.6(5),~~
C_{\epsilon\epsilon}^{\epsilon} A^{-}=139.3(9)
\nonumber
\end{equation}
Thus we see that, in the range of value of $s$ of experimental interest,
the third term in the CPT expansion is of the same size of the second one and cannot be neglected in 
the correlators. Moreover we shall see below, by comparing with Monte Carlo simulations that, 
in the same range, the sum of these two terms almost saturates the correlator, i.e that the higher 
order terms neglected in the above expansion give a contribution to the correlator almost 
negligible (and in any case never larger than $2\%$) with respect to the first three terms. 
They can thus be safely neglected when comparing with the experimental results. On the contrary, 
within the precision of the experimental data, both the second and the third term of 
the CPT expansion are necessary to fit the data.

\section{Monte Carlo simulations}
We compared our results with a set of Monte Carlo simulations of the Ising model using the standard 
nearest neighbour action for which the critical temperature is known with high precision
$\beta_{c}= 0.22165462$ \cite{hasen}.
The temperature perturbation from the critical point is defined through the parameter $t_{lat} 
\equiv \beta_{c} - \beta$. With this convention in the high 
temperature phase we have $t_{lat}>0$. We performed our simulations with a
standard Metropolis updating and multispin coding technique on a cubic lattice with periodic 
boundary conditions. We fixed the lattice size $L=300$, which was the maximum value compatible with 
the computational resources at our disposal. We evaluated the correlators for many distances in 
different simulations, so our data are uncorrelated, and we sampled about $10^{9}$ configurations 
for each simulation, with a thermalization time of $10^{5}$ sweeps.

We defined the lattice discretization of the spin  and energy operators as:
 $$\sigma_{lat} \equiv 1/L^{3} \sum_{i} \sigma_{i},~~~ \epsilon_{lat} \equiv 1/(3L^{3}) \sum_{< ij>} 
\sigma_{i} \sigma_{j} - \epsilon_{an},$$ where $\epsilon_{an}$ is the energy analytic part that 
must be subtracted. Also  $\epsilon_{an}$ is known in the Ising case with high precision:
 $\epsilon_{an} = \epsilon_{cr} + C \: t_{lat} $ , where 
$\epsilon_{cr} = 0.3302022 (5) $ and $C = 9.7 (1)$ are respectively the energy and the 
specific heat at the bulk \cite{hasen}.

We chose $t_{lat}$ small enough so as to have a large enough value of the correlation length
$\xi$, but not too small in order to avoid finite size effects. The optimal choice turned out to be
 $t_{lat} \gtrsim 10^{-4}$, for which $\xi^{+} 
\lesssim 65$ lattice spacings in the high temperature phase and $\xi^{-} \lesssim 34 $ lattice 
spacings for low temperatures. We verified that the finite size effects were negligible within 
our current precision for these temperatures.

\begin{figure}
\vskip -1.cm
\includegraphics[scale=0.4]{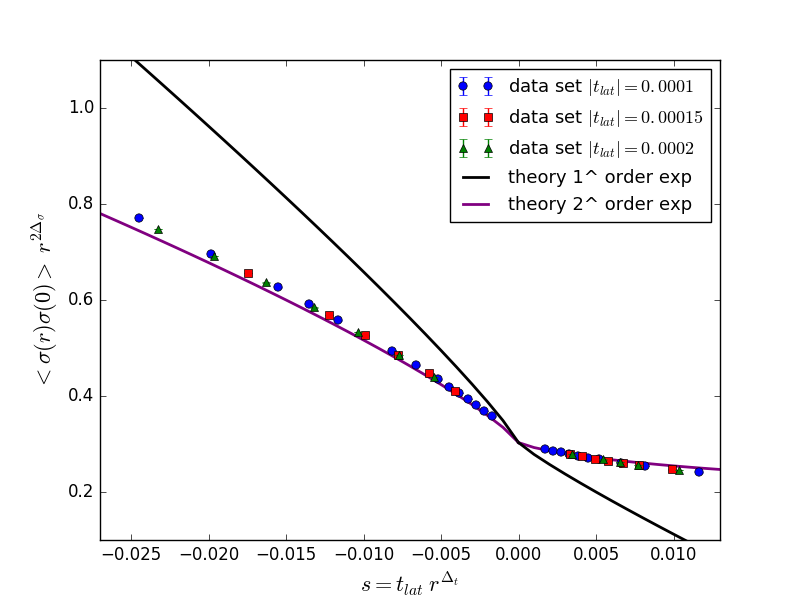}
\caption{\label{fig1} Comparison of the Monte Carlo data with our CPT prediction.}
\end{figure}

The results of our simulations are reported together with the CPT expansion in fig.\ref{fig1}. We also 
report the prediction truncated at the second order. It is easy to see that the universal third 
term of the expansion that we evaluate in this letter gives a large contribution to the correlator, 
which cannot be neglected within the precision of current data  (and also, as we shall see, within 
the precision of experimental estimates), and that our complete CPT expansion agrees remarkably well 
with Monte Carlo data. 

In order to estimate the contribution of the higher order terms neglected in the CPT expansions of 
eq.s(\ref{corr1},\ref{corr2},\ref{corr3}) we 
fitted the spin-spin correlator with eq. (\ref{corr1}), keeping the coefficient of the last term 
(i.e. $\partial_{t} C_{\sigma\sigma}^{1}$) as a free parameter.

The results of the fits are reported in tab.\ref{tab1} where we quoted separately statistical and 
systematic (mainly due to the uncertainty in $A^{\pm}$ and $R_\sigma$) errors.
The results, for all temperatures both above and below $\beta_c$, turn out to be remarkably close to 
the theoretical prediction ${\partial_t C_{\sigma\sigma}^{1}} \simeq 65.7762 $ thus showing that in 
this range of values of the scaling variable $s$ higher order terms in the CPT expansion are almost 
negligible, as also suggested by fig.\ref{fig1}.

\begin{table}[h]
\centering
\begin{tabular}{|l|cccc|}
\hline
$t_{lat}$ & $r_{min}$& $r_{max}$ & $\partial_t C_{\sigma\sigma}^{1}$ & $\chi^{2}/d.o.f$ \\
\hline
$+10^{-4}$  & 6  & 20 & 61.4 (0.9)[1.2] & 0.7 \\
$-10^{-4}$  & 6  & 20 & 60.9 (0.9)[1.5] & 0.8  \\
\hline
$1.5 \cdot 10^{-4}$ & 7 & 14 & 61.3 (0.8)[1.0] & 1.0 \\
$-1.5 \cdot 10^{-4}$ & 8 & 20 & 61.1 (0.9)[1.8] & 1.1 \\
\hline
$2 \cdot 10^{-4}$ & 6 & 13 & 61.0 (0.8)[1.0] & 0.7 \\
$-2 \cdot 10^{-4}$ & 8 & 20 & 61.6 (0.7)[1.5] & 1.2 \\
\hline
\end{tabular}
\caption{ Results of the fits to the spin-spin correlator performed keeping $\partial_t 
C_{\sigma\sigma}^{1}$ as free parameter. 
The columns $r_{min}$, $r_{max}$ indicate the range of distances sampled. Statistical errors are 
reported in round brackets, while the systematic ones, mainly due to the constant $R_{\sigma}$, are 
reported in square brackets. }
\label{tab1}
\end{table}

\section{Scattering Function and Comparison with experimental results}
By Fourier transforming the spin-spin correlator it is easy to construct the Scattering Function 
(for a detailed discussion see for instance \cite{MartinMayor:2002mb})
which turns out to have exactly the form predicted by Fisher and Langer \cite{Fisher1968}
\begin{equation}\label{corr1t}
g(q) = \frac{C_{1}^{\pm}}{q^{2-\eta}} \left( 1 + \frac{C_{2}^{\pm}}{q^{(1-\alpha)/\nu}} + 
\frac{C_{3}^{\pm}}{q^{1/\nu}} \right)
\end{equation}
with $q=k\xi$, where $k$ is the momentum-transfer vector and $\xi$ the correlation length. 
The coefficients of the expansion can be deduced exactly from 
the CPT analysis discussed above:
$$C_{2}^{\pm} 
\xi_{\pm}^{-\Delta_\epsilon} = a_{f} C_{\sigma\sigma}^{\epsilon} A^{\pm} 
|t|^{\frac{\Delta_\epsilon}{\Delta_t}}~,~~~~ 
C_{3}^{\pm} \xi_{\pm}^{-\Delta_t} = b_{f} {\partial_{t} C_{\sigma\sigma}^{1}} t $$
where $a_{f}\simeq -0.668025...$ and $b_{f}\simeq -1.02863...$ are numerical coefficients coming from the Fourier transform of the power 
law terms of eq.(\ref{corr1}). 
Combining all the factors we finally obtain:
$$C_{2}^{+}=2.54(2),~~C_{2}^{-}=-1.86(1),$$
$$C_{3}^{+}=-3.64(1),~~ C_{3}^{-}=1.28(1)~.$$

\begin{figure}\vskip -0.5cm
\includegraphics[scale=0.4]{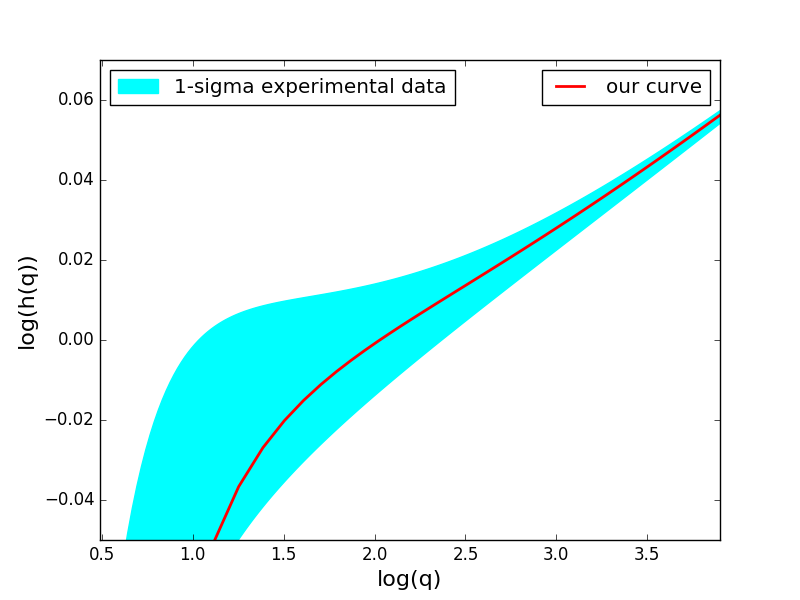}
\caption{\label{fig2} Plot of $h(q)=ln(g(q)/g_{OZ}(q)$. The continuous line is our prediction while 
the light-blue region marks the experimental estimate quoted in \cite{exp}  }
\end{figure}

This result allows a set of interesting theoretical end experimental checks.
First of all they agree remarkably well with the results obtained in\cite{MartinMayor:2002mb} with 
the $\epsilon$-expansion
within the Bray approximation \cite{Bray1976a} (see tab.\ref{tab2}). They also show that the Bray 
approximation which assumes $C_{2}^{+}+ C_{3}^{+}\sim-0.9$ is indeed a good approximation of
the CPT result which gives $C_{2}^{+}+ C_{3}^{+}=-1.10(3)$. This is reassuring, since it shows that 
(as we would expect) the $\epsilon$-expansion actually "knows" that the 3D Ising model is 
conformally invariant.
As expected they also agree with the Fourier Transform of our Monte Carlo results (see the first 
column of tab.\ref{tab2}).
What is more interesting is that they also agree with a set of experimental measures of the 
scattering function obtained from a small-angle neutron scattering experiment on a
sample of ${\bf CO_2}$ at critical density \cite{exp}.
We report our result together with the experimental estimates in fig.\ref{fig2} where we plotted
the scattering function in a log-log scale so as to show the large $q$ scaling behaviour as a 
straight line and normalized it (as usual) to the Ornstein-Zernicke (OZ) function: 
$g_{OZ}=1/(1+q^2)$ so as to evidentiate the large $q$ deviations with respect to the OZ behaviour 
(which describes the  small $q$ behaviour of the scattering function).
We plot our results with a red line and in light blue the region of $\pm\sigma$ uncertainty quoted 
in \cite{exp} as the best fit result of their experimental scattering data
within the Bray approximation. We report their best fit estimates for $C_{3}^{\pm}$, $C_{2}^{\pm}$ 
in the last column of tab.\ref{tab2}, together with the results from our conformal perturbation 
theory, the MC simulations, and the estimates obtained via $\epsilon$-expansion in 
\cite{MartinMayor:2002mb}.

\begin{table}[h]
\begin{tabular}{|c|cccc|}
\hline
            & MC & CPT & $\epsilon$-exp & experim. \\
\hline
$C_{2}^{+}$ &   & 2.54 (2) & 2.56 & 2.05 (80) \\
$C_{2}^{-}$ &   &-1.86 (1) & -1.3 & -1.5 (8) \\
\hline
$C_{3}^{+}$ &  -3.42 (6)  &  -3.64 (1)  & -3.46 & -2.95 (80) \\ 
$C_{3}^{-}$ &   1.20 (2)  &   1.28 (1)  &  0.9  &  1.0 (8) \\
\hline
\end{tabular}
\caption{Values of $C_{2}^{\pm}$, $C_{3}^{\pm}$  obtained with different approaches. The 
column MC shows the result of the fit to our Monte Carlo simulations. The second column 
contains the CPT estimates (for $C_{2}^{\pm}$ we used the amplitudes 
$A_{\pm}$ taken from ref.\cite{hasen} as additional external input). The third column shows the results of 
the $\epsilon$-expansion within the Bray approximation\cite{Bray1976a} from 
\cite{MartinMayor:2002mb}, and the last one reports 
the experimental estimates from \cite{exp}.}

\label{tab2}
\end{table}


\section{Conclusions}
Our main goal in this letter was to extend the knowledge reached in these last years on higher dimensional 
CFTs to the off-critical, scaling, regime of the models. We concentrated in particular on two point 
functions in the 3D Ising model perturbed by the thermal operator, but we see no obstruction to 
extend this program to other universality classes or to three-point functions. An important aspect 
of our analysis was the identification of a set of terms in the CPT expansion which are universal as 
a consequence of the conformal symmetry of the underlying fixed point theory. These universal 
amplitude combinations were already discussed, and compared with two-dimensional data, in 
\cite{Caselle:2003ad}. 
Our results are in good agreement both with Monte Carlo simulations and with a set of experimental 
results in systems belonging to the 3D Ising universality class. We think that in future these 
techniques will allow us to vastly extend our ability to describe experimental data in the scaling 
regime of three dimensional critical points.

\vskip 0.2cm
{\bf Acknowledgments:}
We would like to thank A. Amoretti, F. Gliozzi, M. Panero, A Pelissetto and E. Vicari for useful discussions and suggestions and the 
INFN Pisa GRID data center for supporting the numerical simulations. 

\vskip -0.5cm
\bibliography{CFT}
\end{document}